# A thermal bonding method for manufacturing Micromegas detectors

Jianxin Feng[a,b], Zhiyong Zhang[a,b]*, Jianbei Liu[a,b]*, Binbin Qi[a,b], Anqi Wang[a,b], Ming Shao[a,b], Yi Zhou[a,b]

[a] *State Key Laboratory of Particle Detection and Electronics, University of Science and Technology of China, Hefei 230026, China*

[b] *Department of Modern Physics, University of Science and Technology of China, Hefei 230026, China*

**Abstract**

For manufacturing Micromegas detectors, the "bulk" method based on photoetching, was successfully developed and widely used in nuclear and particle physics experiments. However, the complexity of the method requires a considerable number of advanced instruments and processing, limiting the accessibility of this method for production of Micromegas detectors. In view of these limitations with the bulk method, a new method based on thermal bonding technique (TBM) has been developed to manufacture Micromegas detectors in a much simplified and efficient way without etching. This paper describes the TBM in detail and presents performance of the Micromegas detectors built with the TBM. The effectiveness of this method was investigated by testing Micromegas detector prototypes built with the method. Both X-rays and electron beams were used to characterize the prototypes in a gas mixture of argon and $CO_2$ (7%). A typical energy resolution of ~16% (full width at half maximum, FWHM) and an absolute gain greater than $10^4$ were obtained with 5.9 keV X-rays. Detection efficiency greater than 98% and a spatial resolution of ~65 μm were achieved using a 5 GeV electron beam at the DESY test-beam facility. The gas gain of a Micromegas detector could reach up to $10^5$ with a uniformity of better than 10% when the size of the avalanche gap was optimized thanks to the flexibility of the TBM in defining the gap. Additionally, the TBM facilitates the exploration of new detector structures based on Micromegas owing to the much-simplified operation with the method.

## 1. Introduction

The Micromegas detector is a typical micropattern gaseous detector (MPGD) that was invented in 1996 [1]. It has a two-stage structure: a 3–5 mm drift region for primary ionization of charged particles and an avalanche gap of about 0.1 mm. Owing to its excellent energy resolution, rate capability, and possibility for large area, several methods have been developed to manufacture this type of detector. Producing the

*Corresponding author*: zhzhy@ustc.edu.cn, liujianb@ustc.edu.cn

narrow avalanche gap is the most challenging part of the fabrication process of a Micromegas detector.

Among the different methods of manufacturing Micromegas detectors, the bulk method based on photoetching [2-3] has been well developed and substantiated through its application at various experiments, such as T2K [4], COMPASS RICH [5], and the ongoing upgrade of the ATLAS NSW [6]. Micromegas detectors built with the bulk method have exhibited excellent performance, such as gas gain of $2\times10^4$ [2], energy resolution below 20% in argon based mixtures and gain uniformity of 3.6% [7]. However, this method makes pillars that support the avalanche gap by etching photopolymer. So, it heavily depends on the equipment for performing exposure, etching and lamination, and professional operation of advanced devices. To avoid the etching operation, a new method based on a thermal bonding process for manufacturing Micromegas detectors as a result of decade long effort [8-9]. This method has shown the capability of producing Micromegas detectors and its derivatives in an easier and more efficient way by using simple devices and conventional materials.

This paper is organised as follows. The TBM is described in detail in section 2, including its basic principle, fabrication process and advantages. In section 3, performance of Micromegas detectors fabricated with the TBM was intensively studied. The effectiveness of the TBM is demonstrated by the results from the study including energy resolution, avalanche gain, spatial resolution, detection efficiency, and gain uniformity. Moreover, Micromegas detectors built with the TBM have been successfully used in different experiments, for instance, as trackers for muon tomography, neutron beam monitors, and double micro-mesh gaseous detectors (DMM) for single electron detection. These application examples are briefly discussed in section 4. Finally, optimization of the avalanche gap of a Micromegas detector with the TBM is presented in section 5. Significant improvements in gas gain and its uniformity are achieved with the optimization.

## 2. Thermal bonding method (TBM)

*2.1 Principle of the method*

In this method, the avalanche structure of the Micromegas detector is manufactured by bonding the stretched stainless-steel mesh directly onto the readout printed circuit board (PCB) using a hot rolling machine. Figure 1 shows a schematic of the method. As opposed to etching pillars, the TBM uses two hot rollers to melt and press the spacers made of a thermal bonding film, which have an adhesive–polyester–adhesive triple-layer structure [10], to support the stainless-steel mesh [11] onto the readout anode. Spacer adhesives become solid on cooling and fix the mesh on it, thus forming the micro-gap between them. This whole process involves no chemical agents, hence an etching-free feature of the thermal bonding method. The details of the method will be described in section 2.2.

The spacers with a diameter $\leqq 1$ mm are arranged at a distance of ~10 mm on the readout PCB, as depicted in Figure 2. This special arrangement ensures that the dead area, which is defined as the ratio of the spacer to the active area, is constrained to a small fraction (< 1%), similar to the bulk method where pillars with a diameter of 0.2–0.4 mm are set at a distance of 2 mm from each other [2]. Particularly, the large spacer pitch facilitates cleaning of the avalanche region and is prominent to evade sparking in high electric fields. Therefore, a much higher tensile force is required on the mesh to resist the electrostatic force between the mesh and anode.

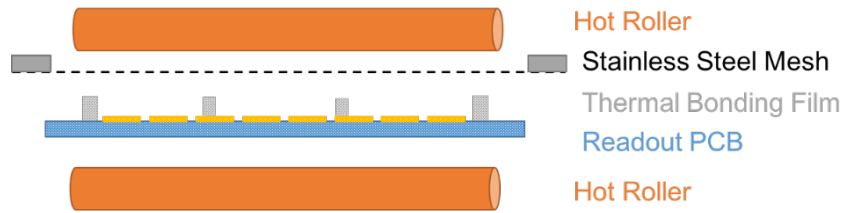

Figure 1: Schematic of the TBM for Micromegas.

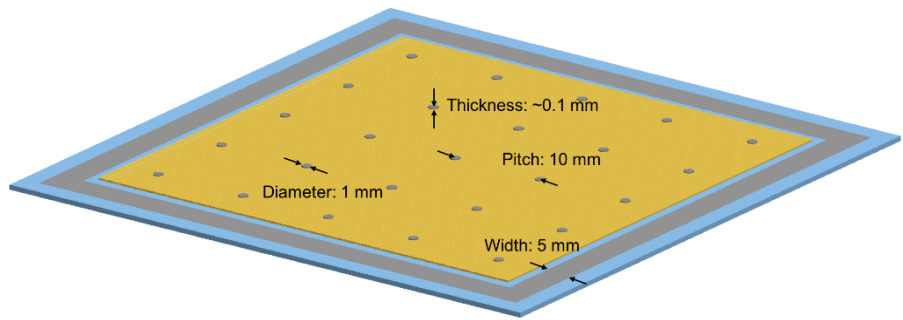

Figure 2: Schematic design of the spacer arrangement.

## 2.2 Implementation of the manufacturing process

Following the principle mentioned in subsection 2.1, the manufacturing process can be implemented step-by-step as shown in Figure 3 and demonstrated below:

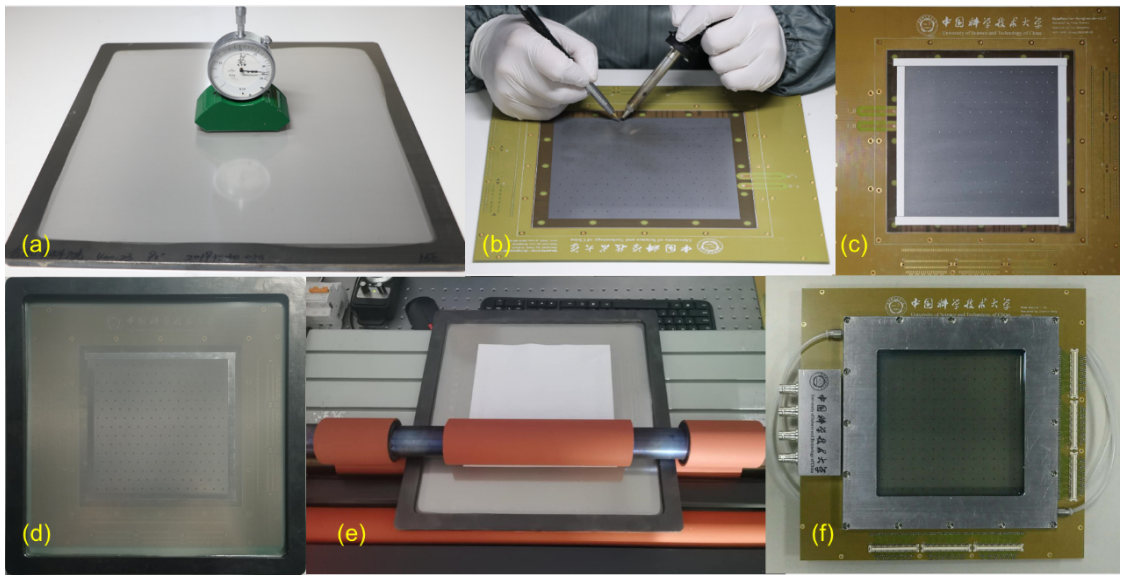

Figure 3: Fabrication process for Micromegas using the TBM. (a) The stainless-steel woven mesh is stretched over a tension of 25 N/cm and glued on a metal frame. (b) A manual method for pre-

setting the spacers on the readout PCB anode. (c) The view after setting borders (the thermal bonding film in white colour, 3–6 mm in width), which fix the mesh and define the active area for the Micromegas. (d) Stacking the stretched mesh on the PCBs for thermal bonding. (e) Manufacturing the avalanche structure of the Micromegas by thermal bonding using a hot roller. (f) Completed view of the thermal bonding Micromegas after assembling the drift electrode, gas chambers, connectors, and other associated components.

*2.2.1 Preparations:* The preparation work of the TBM primarily includes stretching of the stainless-steel mesh and cutting the spacers. The mesh is fully stretched to a high tension of more than 25 N/cm (depending on the spacer settings) with a commercial stretching machine, as shown in Figure 3 (a). The small-sized spacers are produced by laser cutting, which is a conventional method for micro- and nanomanufacturing. To avoid melting the adhesive on the film while cutting, the pulse width of the laser machine must be less than 1 ns.

*2.2.2 Resistive anode*: A germanium layer (grey) coated on the surface of the PCB, as shown in (Figure 3 (b) and (c)), is used as a resistive anode. Its sheet resistivity varies from tens of MΩ/sq to hundreds of MΩ/sq corresponding to a film thickness of 500–100 nm.

*2.2.3 Pre-setting for the spacers and borders*: The spacers and borders, made of a thermal bonding film, are set on the readout PCB in advance by pre-heating. As shown in Figure 3 (b), the spacer is picked and pressed on the PCB anode by using a needle (in right hand). Further, a hot air pencil, controlled by the left hand, is used to flush the spacer with hot air for a few seconds, so that the adhesive on the spacer will melt slightly and attach onto the PCB. Moreover, the borders (white) are set following a similar operation, and the completed subsystem is shown in Figure 3 (c).

*2.2.4 Thermal bonding process*: As shown in Figure 3(d), the tensile mesh is placed directly on the PCB where the spacer and border have been set. They are inserted together between two hot rollers (orange) and driven by them to move back and forth for several rounds at an empirical temperature ~150 ℃. Therefore, the thermal bonding adhesive on the spacer and border is melted and the mesh is fixed onto the readout PCB after cooling to room temperature (20–30 °C).

*2.2.5 Assembly*: After the thermal bonding process, the Micromegas detector is finally assembled by trimming the mesh extending out of the sensitive area, installing the drift electrode and gas chamber, soldering the readout connectors, and other associated processes. Figure 3 (f) shows an image of assembled detector, which has a sensitive area of $150 \times 150$ mm$^2$.

*2.3 Advantages of the TBM*

It is obvious from the above implementation of the method, the TBM avoids using any environmentally harmful chemical agents, which benefits the long-term development and application of the method. The simple process and minimal requirement of equipment establishes it as an economical and easy to operate process in the laboratory, while simultaneously maintaining good performance of the fabricated detectors.

Additionally, the thickness of the avalanche gap is determined by the thermal bonding film, and is therefore, convenient to adjust the gap for optimization and specific applications. The commercially available thermal bonding films have different thicknesses, ranging from several tens of microns to hundreds of microns, allowing the fabrication of avalanche gap as minimum as ~30 μm.

However, it is worth noting that, the avalanche gap is related to the conditions such as temperature and pressure in the thermal-bonding process as well. Our studies show that the adhesive layers of the spacer should be fully squeezed while heating to press the spacer to a thickness that is identical to the unmelting middle layer. This helps to achieve maximum uniformity of the TBM. The impact of this effect will be discussed in section 5.

### 3. Performance of TBM Micromegas detectors

To ascertain the feasibility of the TBM, several Micromegas prototypes with a sensitive area of 150 × 150 mm$^2$ were fabricated and tested (MM01–MM05). The mesh has 400 lines per inch (LPI) with 19 μm wire diameter and 49% opening rate. A 250 μm thick thermal bonding film, configured as a stack of 75–100–75 μm, was used to develop the spacers. An avalanche gap of ~110 μm was measured with a microscope in the cross-section of a sawed detector after solidification of the glue that had been filled into the gap. The prototypes were designed as two-dimensional (2D) readout PCB with strips each having a 400 μm pitch. As shown in Figure 4, two orthogonal layers of strips were set in the inner layer of the readout PCB. The widths of the strips were 80 μm and 320 μm for the upper (X direction) and lower (Y direction) layers, respectively. The drift gaps between the drift cathode and the mesh were 5 mm and a gas mixture of argon and $CO_2$ (7%) was used as the detector active medium.

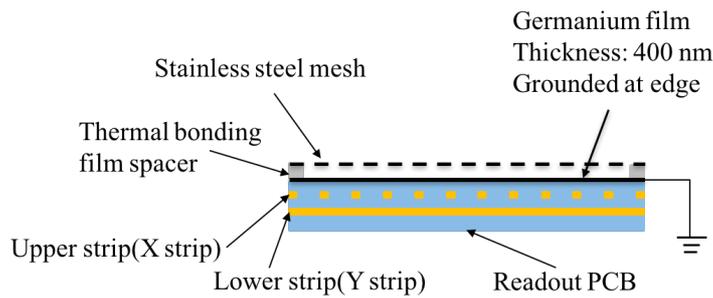

Figure 4: Schematic representation of the Micromegas prototype structure with a 2D readout.

*3.1 Testing with an X-ray source*

The basic performance of the prototypes was assessed in the laboratory, using a gas mixture of argon and $CO_2$ (7%), with the $^{55}$Fe radioactive source that emits 5.9 keV X-rays. An energy resolution of ~16% (FWHM) (shown in Figure 5) and a high gain (>10$^4$) (shown in Figure 6), were the typical performance of these prototypes. The energy resolution was extracted by fitting the energy spectrum with a sum of three Gaussian functions, named $K_{alpha}$, $K_{beta}$ of $^{55}Mn$ and an escape peak of argon.

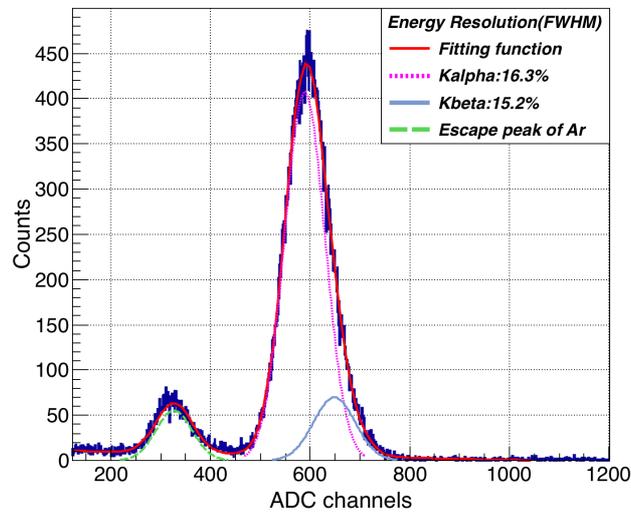

Figure 5: X-ray spectrum of the $^{55}$Fe source at a gain of ~5000.

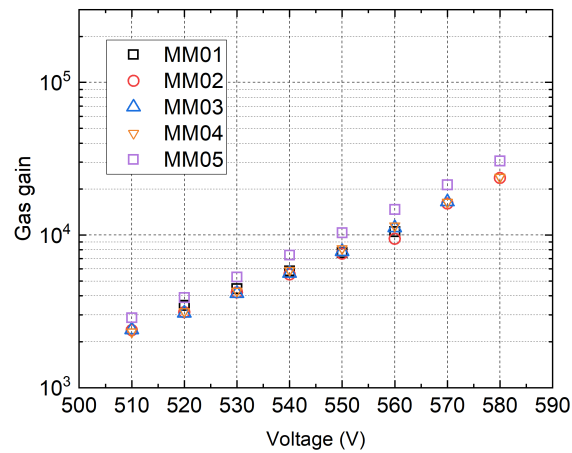

Figure 6: Dependence of gain on mesh voltage.

The uniformity of gain is defined as the ratio of the variance to the mean of the peak values of the energy spectra and was tested for every 20 × 20 mm$^2$ block on the detector sensitive area. As shown in Figure 7, the dots represent the positions of X-ray irradiation for scanning the uniformity of the detector gain. The scanned uniformity was typically ~16%, as shown in Figure 8.

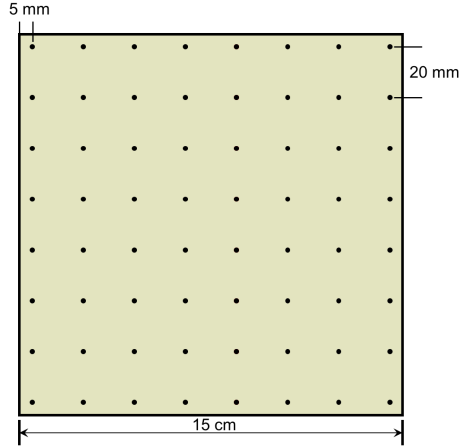

Figure 7: Schematic diagram showing the scanning positions (dots) in the sensitive area of the prototypes.

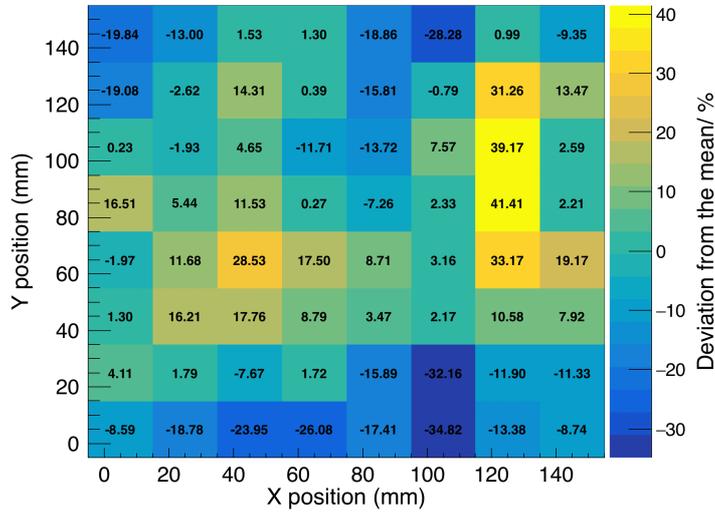

Figure 8: Map the uniformity of the gain of one of the prototypes (~16% at mean gas gain of 19082.2 and R.M.S. of 3083.5).

*3.2 Testing with electron beam*

The detection efficiency and spatial resolution of the Micromegas prototypes were investigated with the 5 GeV electron beam at Deutsches Elektronen–Synchrotron (DESY), Germany [12]. Three prototypes (MM01, MM02, and MM03) were installed in the beam line, as shown in Figure 9. A dedicated front-end electronics (FEE) system for the MPGDs, based on the AGET ASIC chip [13-14], was developed to read out the strip signals. Each channel can record 512 sampling points with adjustable sampling frequency from 3 MHz to 100 MHz and optional peaking time from 50 ns to 1 μs. However, 40 MHz sampling frequency and 1 μs peaking time were selected to record the data in this test.

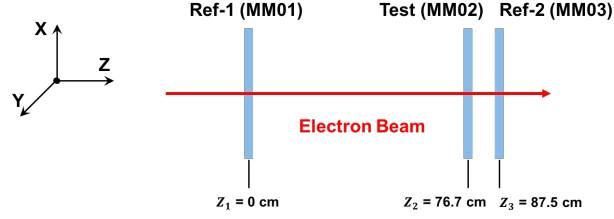

Figure 9: Electron beam test setup.

The electrons behave as the minimum ionization particles (MIPs) in the 5 mm drift region of the detectors. The distribution of the energy deposition measured as the sum of ADCs of the fired strips, was fitted with a Landau function convoluted with a Gaussian function accounting for the detector energy resolution. One example of the measured energy distribution along with the fitting is illustrated in Figure 10 for MM02. The mesh voltage was 540 V and the gas gain was ~ $2 \times 10^4$ for this prototype.

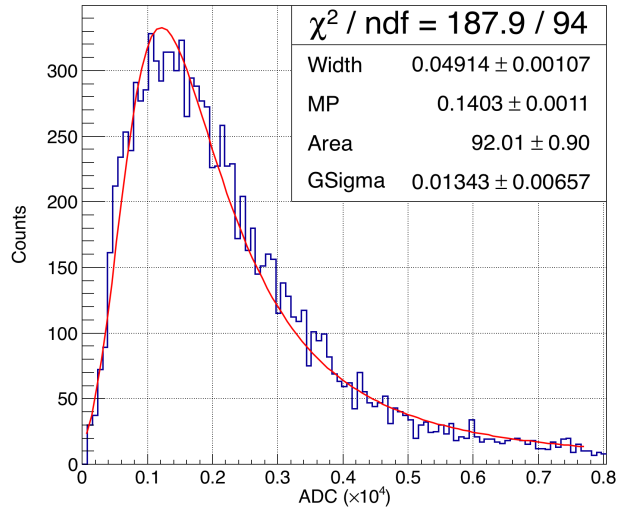

Figure 10: Detector response to the electron beam.

Further, to estimate the spatial resolution, MM01 and MM03 were used as reference tracking detectors and MM02 was the test detector. The deviations of the tested MM02 ($\Delta X$, here take the deviation of X strip as an example) were defined as the differences between the hit positions ($X_t$) recorded by MM02 and the expected positions ($X_{exp}$) derived from the reference trackers. The $X_{exp}$ can be simply calculated by $a \cdot Z_t + b$, where $Z_t$ is the installing position of MM02 in the Z direction, and $a$ and $b$ are the slope and intercept of the reference tracker, respectively.

The spatial resolution can be estimated by the following equation:

$$\delta^2(\Delta X) = (1 + \left(\frac{Z_2 - Z_t}{Z_2 - Z_1}\right)^2 + \left(\frac{Z_1 - Z_t}{Z_2 - Z_1}\right)^2)\delta^2(X_t) \quad (1)$$

where, $\Delta X = X_t - X_{exp}$ is the residual, $\delta(X_t)$ is the square root of the variance of the residual represents the spatial resolution of the (MM02) detector under test. For the electron beam test, the detectors were placed at $Z_1 = 0$ cm, $Z_t = 76.7$ cm, and $Z_2 = 87.5$ cm, respectively, as shown in Figure 9. Thus,

$$\delta(\Delta X) = 1.36\delta(X_t) \qquad (2)$$

The $\delta(\Delta X)$ is obtained by fitting the residual distribution with a double Gaussian function, as shown in Figure 11. Angular divergence of the electron beam and multiple scattering of the electrons both contribute to the tail events in the residual distribution, which are accounted for by the wide component of the double Gaussian function (dotted line). The narrow component or the core Gaussian distribution (dashed line) has a standard deviation of 88.7 ± 1.42 μm and hence a spatial resolution $\delta(X_t)$ of 65.22 ± 1.04 μm.

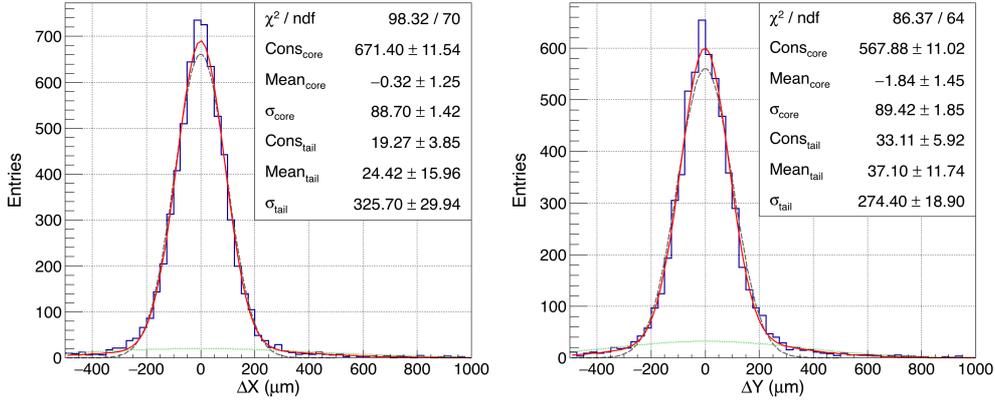

Figure 11: Deviation distributions of the Micromegas prototypes in the X (left) and Y (right) directions.

The detection efficiency is defined as the ratio of effective events recorded by the test prototype (MM02) to coincidence events of reference trackers (MM01 and MM03), where the effective events of MM02 were selected with the condition of $\Delta X < 10 \cdot \delta(\Delta X)$. Figure 12 shows the detection efficiency of the individual X and Y strips. The squares represent the efficiency of the X strips, and the inverted triangles represent the efficiency of the Y strips. Efficiencies of both strips reach higher than 98% on the plateau area when the mesh voltage exceed 530 V, which corresponds to a gas gain of ~5000.

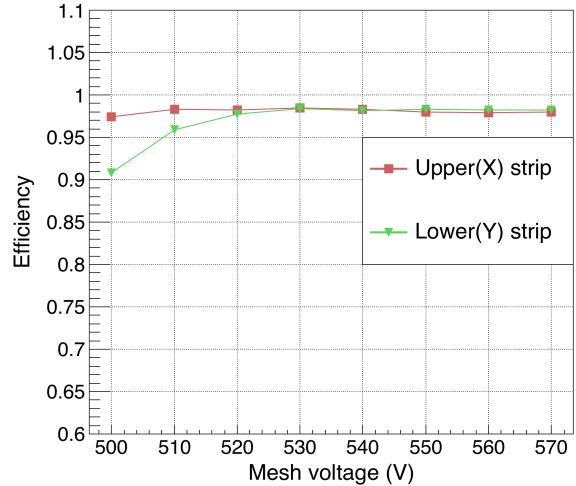

Figure 12: Detection efficiency of thermal bonding Micromegas.

## 4. Applications of TBM Micromegas detectors

The best way to ascertain the stability and performance of the detectors is to use them in experiments. Several Micromegas detectors have been designed and manufactured using the TBM for different applications. A six-layer tracker system was built as a cosmic ray telescope using Micromegas detectors. In this telescope, each detector has a sensitive area of 90 × 90 mm$^2$ and has been operated periodically for approximately two years, which indicates good stability of these detectors [15].

A 2D readout Micromegas-based neutron detector has been employed as beam monitor in the China Spallation Neutron Source (CSNS) facility, which is an advanced accelerator-based neutron facility proposed to serve both scientific research and industrial applications [16]. Using this novel detector, a 2D neutron beam spot distribution at the Back-n beam line was obtained for the first time [17].

Additionally, owing to the flexibility of the TBM, a new type of detector, having high-gain and low ion-backflow (IBF), with a double micro-mesh gaseous structure (DMM) has been developed using TBM for single photoelectron detection [18-19]. A schematic diagram of the DMM structure is shown in Figure 13. In this, two stacked meshes are bonded to produce a two-stage cascade avalanche structure. The observed high gain (>10$^6$) and low IBF ratio (~0.05%) of the DMM indicate that it can be potentially used in gaseous photomultiplier tubes (Gas-PMTs) and other applications, for instance, readout of the time projection chamber detectors for the future collider experiments [20-21].

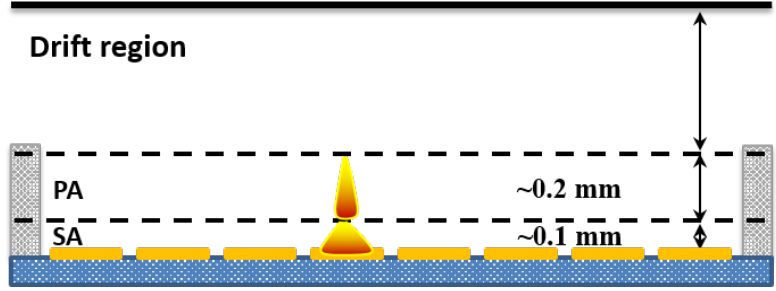

Figure 13: Schematic diagram of DMM.

## 5. Micromegas detector optimization with TBM

Following the idea mentioned in subsection 2.3, the temperature and pressure for thermal bonding were increased to completely squeeze the spacers and five optimized prototypes (OptimizedMM01-05) were fabricated and subsequently tested. The avalanche gaps were measured to be ~100 μm, which corresponded to the thickness of the polyester layer of the 75–100–75 μm spacers. Consequently, a higher gas gain was obtained compared with the previously obtained gain for the avalanche gap of ~110 μm. As shown in Figure 14, the gas gains of all the optimized Micromegas prototypes can easily exceed $5\times10^4$ at a working voltage above 560 V, and some of them can even reach a gain as high as $10^5$ without breakdown.

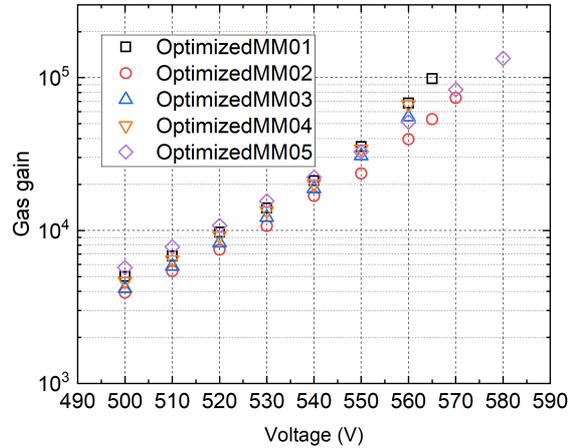

Figure 14: Gas gains of five optimized Micromegas prototypes. The vertical scale is drawn identical to Figure 6 for comparison.

To further evaluate the impact of changing the avalanche gap length from ~110 μm to ~100 μm, a simulation was performed to study the correlation between the gas gain and avalanche gap. The gas gain can be calculated using the following formula:

$$M = e^{\int [\alpha(x)-\eta(x)]dx} \qquad (3)$$

where $\alpha(x)$ and $\eta(x)$ represent the Townsend and attachment coefficients along the moving path $dx$. The Townsend and attachment coefficients for the gas mixture of $Ar/CO_2$ (93%/7%) 1 atm were computed using the Magboltz program [22] for various electric field strengths. The gain variation with the avalanche gap sizes at different mesh voltages is displayed in Figure 15. This result shows that the detector with ~100 μm avalanche gap has higher gain than that with ~110 μm gap at the same mesh voltage. The optimized prototypes (with ~100 μm gap) attained a high mesh voltage, similar to that of the previous prototypes (with ~110 μm) before undergoing breakdown, and thus, achieved a higher gain.

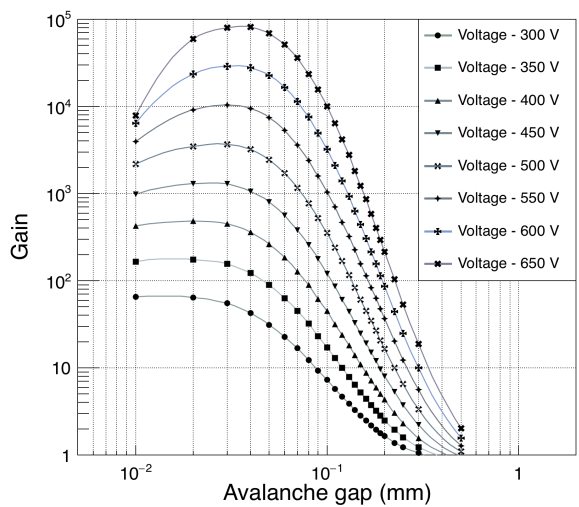

Figure 15: Gain distribution variation along with the avalanche gap size.

Concurrently, squeezing the spacers to their unmelting layer helps to improve the uniformity of the avalanche gap, which leads to an expected significant improvement in the uniformity of the gain. As shown in Figure 16, the value of 8.1% was obtained as compared to the previous result of 16% under the same gain conditions. Additionally, the non-uniformities at different gas gain of ~20000 (Figure 16), ~5000, and ~1000 (Figure 17) were investigated for the OptimizedMM01, and the values of 8.1%, 6.3%, and 5.5% were obtained, respectively. The uniformity within one detector inherently results from the variation of the avalanche gap. Further, the relative differences at low gains become small, which leads to the refinement of the detector uniformity.

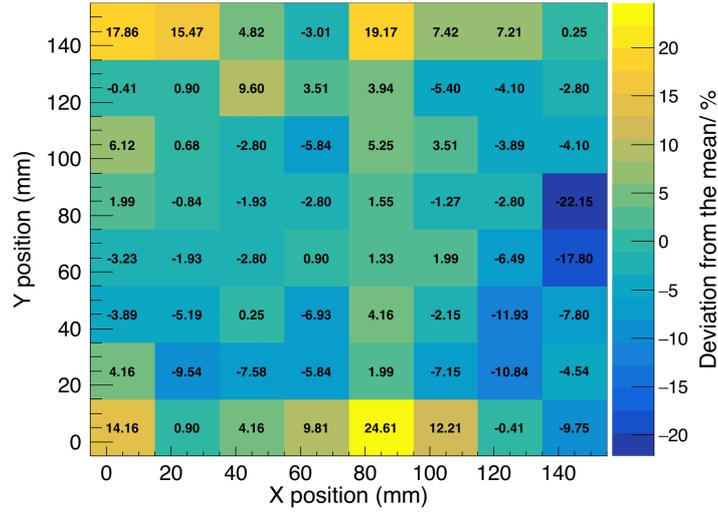

Figure 16: Map the uniformity of gain of the prototypes (OptimizedMM01, 8.1% at mean gas gain of 19950.1 and R.M.S. of 1610.5).

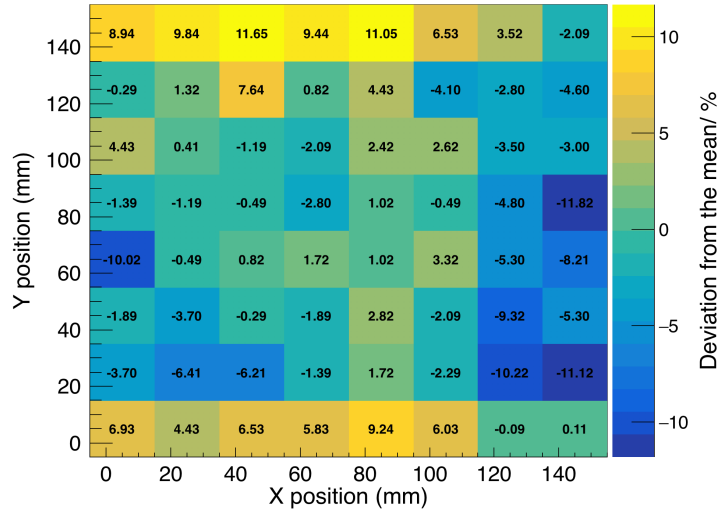

Figure 17: Map the uniformity of prototypes (OptimizedMM01, 5.5% at mean gas gain of 996.9 and R.M.S. of 54.3).

## 6. Conclusions

The TBM, an innovative method for Micromegas fabrication, has been proposed and extensively developed in the past decade. In this TBM, a large distance between the spacers makes it easier to keep the avalanche region clean and reduce the sparking probability. Good performance characteristic has been achieved in terms of gas gain, energy resolution, spatial resolution, and detection efficiency with the Micromegas prototypes. A high gain of ~$10^5$, an energy resolution of 16%, and a uniformity of ~ 6.3% (at gain of ~10000) were obtained with 5.9 keV X-rays as the ionizing radiation. Moreover, better than 98% detection efficiency and a spatial resolution of ~65 μm were observed with a 5 GeV electron beam. The excellent stability of the Micromegas detectors built with the TBM and the capability of the method to make new detector structures have been demonstrated in various applications, such as tracker systems, neutron monitors, etc.

However, in the TBM, manual setting of the spacers is the most time-consuming step. An alternative method of automatically setting spacers is being developed, which is critical for the mass production of large area detectors.

**Acknowledgements**

The authors wish to thank the Hefei Comprehensive National Science Center for its strong support. This work was partially performed at the University of Science and Technology of China (USTC) Center for Micro and Nanoscale Research and Fabrication, and we thank Yu Wei for his help in the nanofabrication steps for Ge coating, and Dianfa Zhou for his help on the laser cutting of thermal spacers. We extend our gratitude to Mr. Jacob Ma for his efforts in the fabrication and testing of the tracking Micromegas prototypes. The measurements leading to the results reported in this paper were performed at the test beam facility at DESY, Hamburg (Germany), a member of the Helmholtz Association (HGF). Meanwhile, we sincerely appreciate the assistance of Qian Liu from UCAS, Baolin Hou and Zhen Chen from USTC during the DESY beam test.

**Funding.** This work was supported by the National Key Programme for S&T Research and Development [grant number 2016YFA0400400]; Program of National Natural Science Foundation of China [grant numbers 11605197, 11935014]; Double First-Class university project foundation of the USTC; and Fundamental Research Funds for the Central Universities.